\def\thorn{I\kern-0.4em\raise0.35ex\hbox{\it o}}

\documentclass[twocolumn,showpacs,preprintnumbers]{revtex4}
\usepackage{graphicx}
\usepackage{amssymb}

\begin{document}
\title{Gravitational radiation from dynamical black holes}
\author{Sean A. Hayward}
\affiliation{Institute for Gravitational Physics and Geometry, The Pennsylvania
State University, University Park, PA 16802, U.S.A.}
\date{Revised 14th November 2005}

\begin{abstract}
An effective energy tensor for gravitational radiation is identified for
uniformly expanding flows of the Hawking mass-energy. It appears in an energy
conservation law expressing the change in mass due to the energy densities of
matter and gravitational radiation, with respect to a Killing-like vector
encoding a preferred flow of time outside a black hole. In a spin-coefficient
formulation, the components of the effective energy tensor can be understood as
the energy densities of ingoing and outgoing, transverse and longitudinal
gravitational radiation. By anchoring the flow to the trapping horizon of a
black hole in a given sequence of spatial hypersurfaces, there is a locally
unique flow and a measure of gravitational radiation in the strong-field
regime.
\end{abstract}
 \maketitle

{\em Introduction.} Gravitational radiation is a prediction of Einstein
gravity, whereby relatively accelerating masses produce ripples in space-time
which propagate outwards with the speed of light. The waves predicted to reach
Earth from distant astrophysical sources are so weak that they have not yet
been directly observed, though strong indirect evidence comes from observations
of the Hulse-Taylor pulsar, for which the orbital period is decaying at the
predicted rate due to loss of energy by gravitational radiation \cite{WT}. To
make direct observations, a new generation of gravitational-wave detectors is
being developed, promising to open a new window on the universe.

A theoretical problem has been the difficulty to define gravitational radiation
in general. In the short-wave approximation for linearized gravitational waves,
one separates space-time ripples from background curvature by assuming that
appropriate length scales exist, leading to an averaged effective energy tensor
\cite{Is}. At infinity in an asymptotically flat space-time, an exact
characterization was found by Bondi and others \cite{TGR}, with the energy flux
of gravitational radiation contributing to an energy-loss equation, whereby the
total mass or energy of the space-time decreases due to outgoing radiation.
Despite these successes, they reflect weak-field definitions. In particular,
since one of the promising sources is the inspiral and merger of a binary
system of black holes or neutron stars, there has been an intense effort to
predict waveforms using numerical simulations. Such simulations cover the
strong-field regime of General Relativity, while waveforms have had to be
extracted subsequently.

This Letter reports an effective energy tensor for gravitational radiation
which can be applied in the strong-field regime, entering an energy
conservation law in the same way as the matter energy tensor. At infinity, they
respectively recover the Bondi energy flux and energy-loss equation, while on a
black-hole horizon, they have the same form as the effective energy tensor and
energy conservation equation recently found for dynamical black holes
\cite{bhd}, the latter originally derived by Ashtekar \& Krishnan \cite{AK}.
The measure of gravitational mass-energy enclosed by a spatial surface is taken
to be the Hawking energy \cite{Haw}, which was useful in both cases above,
reducing respectively to the Bondi mass and irreducible mass. Attempts to find
similar results for other definitions of energy have not led to justifiable
physical interpretation.

The main intended scenario is one of the above simulations, where a family of
spatial hypersurfaces has been numerically evolved in time, and one or more
marginal surfaces have been found in each hypersurface, practically locating
the black hole(s). A marginal surface is to be propagated outwards in the
hypersurface, generating a one-parameter family of surfaces, such that the
expansion (of the surfaces in the hypersurface) is constant on each surface. It
transpires that the dual flow, explained below, should also have constant
expansion. These flows were previously described as uniformly expanding
\cite{mon} and generalize inverse mean-curvature flows \cite{PI}. Then the
energy is found to be monotonically increasing along the flow, as described
recently by Mars \cite{Mar}, and to satisfy an energy conservation equation as
described. Further details and developments are to be presented elsewhere
\cite{HMS}.

{\em Spin-coefficient formalism.} It is convenient to use the spin-coefficient
formalism \cite{PR}, since it is easier to understand the interpretation of the
components as energy densities of ingoing and outgoing, transverse and
longitudinal gravitational radiation, by comparing with the Szekeres
characterization \cite{GR} in terms of the Weyl spinor. Since the
spin-coefficient formalism is not generally familiar, a brief summary is given
below, referring to Penrose \& Rindler \cite{PR} for details. Henceforth
equation numbers in triples will indicate equations in that reference. It is
necessary to generalize some expressions and convenient to modify some
notation.

The basic geometrical object is a spin-metric, an antisymmetric 2-form
$\varepsilon_{AB}$ (2.5.2) acting on complex 2-spinors (spin-vectors). It can
be regarded as a square root of the space-time metric, which is expressed as
$g_{ab}=\varepsilon_{AB}\bar\varepsilon_{A'B'}$, (3.1.9), where the bar denotes
the complex conjugate and the abstract indices indicate how a tensor can be
expressed as a spinor dyad. A spin-basis is an ordered pair of non-parallel
spin-vectors $(o^A,\iota^A)$, which defines a null tetrad, a vector basis
(3.1.14): $(l^a,m^a,\bar m^a,l'^a)=(o^A\bar
o^{A'},o^A\bar\iota^{A'},\iota^A\bar o^{A'},\iota^A\bar\iota^{A'})$. Then
$(l,l')$ are real null vectors, whereas $m$ is a complex vector encoding
transverse spatial vectors. In terms of the complex normalization factor
$\chi=\varepsilon_{AB}o^A\iota^B$ (2.5.46), one finds $g(l,l')=-g(m,\bar
m)=\chi\bar\chi$ with the other eight independent (symmetric) contractions of
the tetrad vectors vanishing. Here the metric convention is $({+}{-}{-}{-})$.

The inverse metric is $g^{-1}=2(l\otimes l'-m\otimes\bar m)/\chi\bar\chi$,
where $\otimes$ denotes the symmetric tensor product. The dual basis of 1-forms
(4.13.32) will be denoted by $(n,w,\bar w,n')=g(l',-\bar m,-m,l)/\chi\bar\chi$,
so that $n(l)=w(m)=\bar w(\bar m)=n'(l')=1$, with the other twelve such
contractions vanishing. Then the metric is $g=2\chi\bar\chi(n\otimes
n'-w\otimes\bar w)$. The tetrad covariant derivative operators are denoted by
$(D,\delta,\delta',D')=(\nabla_l,\nabla_m,\nabla_{\bar m},\nabla_{l'})$
(4.5.23). The complex spin-coefficients
$(\kappa,\sigma,\rho,\tau,\varepsilon,\beta,\alpha,\gamma)$ and
$(\kappa',\sigma',\rho',\tau',\varepsilon',\beta',\alpha',\gamma')$ are defined
by (4.5.21) and encode the Ricci rotation coefficients. Further definitions are
of weighted derivative operators $(\thorn,\eth,\eth',\thorn')$ (4.12.15), a
spinor $\Phi_{ABA'B'}$ and a scalar $\Lambda$ encoding the Ricci or Einstein
tensor (4.6.24), the Weyl spinor $\Psi_{ABA'B'}$ (4.6.35) encoding the Weyl
tensor, and its five complex components $\Psi_0,\ldots\Psi_4$ (4.11.9).

{\em Uniformly expanding flows.} It is convenient to choose null coordinates
$(x,x')$ such that $(n,n')=(dx,dx')$, which implies the dual-null gauge
conditions \cite{bhs}
$0=\kappa=\rho-\bar\rho=\varepsilon+\bar\varepsilon=\tau-\bar\alpha-\beta=
\kappa'=\rho'-\bar\rho'=\varepsilon'+\bar\varepsilon'=\tau'-\bar\alpha'-\beta'$.
Then the formalism describes two families of null hypersurfaces labelled by $x$
and $x'$, intersecting in a two-parameter family of transverse spatial
surfaces. The main application is to a spatial hypersurface foliated by
transverse surfaces, which generates a locally unique dual-null foliation,
intersecting in the transverse surfaces.

Consider a one-parameter family of transverse surfaces with generating vector
$\xi$, normal to the transverse surfaces. Then $\xi=\xi^0l+\xi^1l'$, where the
components $\xi^a$ are constant on the transverse surfaces. It is useful to
introduce its dual vector $\xi^*=\hat{*}\xi$ in the normal space \cite{Mar},
where $\hat{*}$ is the vectorial Hodge operator of the normal space, with
orientation chosen so that $\xi^*=\xi^0l-\xi^1l'$. Since the expansions along
$(l,l')$ are $-2(\rho,\rho')$, the condition for the flow to be uniformly
expanding is that $(\rho,\rho')$ are constant on the transverse surfaces.

The flows can also be expressed in terms of the unit-expansion vector (inverse
mean-curvature vector) \cite{Fra}
\begin{equation}
\eta=-H/g(H,H)\qquad H=-g^{-1}(d\log({*}1))
\end{equation}
and its dual $\eta^*=\hat{*}\eta$, where $H$ is the expansion vector
(mean-curvature vector) \cite{Tod}, the signs reflect the metric convention and
${*}1$ denotes the area form of a transverse surface. Then $H=2(\rho'l+\rho
l')/\chi\bar\chi$, $\eta=-(l/\rho+l'/\rho')/4$, $\eta^*=-(l/\rho-l'/\rho')/4$
and the expansions along $(\eta,\eta^*)$ are found to be $(1,0)$. Thus
$\xi=a(\eta+c\eta^*)$ generate all uniformly expanding flows, where $(a,c)$ are
constant on the transverse surfaces \cite{Mar}. This freedom suggests that such
a flow should locally exist for any untrapped or mean convex surface
\cite{mon}, $\rho\rho'<0$, in any hypersurface, though no proof exists.

{\em Energy propagation.} It is convenient to use the area radius
\begin{equation}
R=\sqrt{A/4\pi}\qquad A=\oint{*}1
\end{equation}
where $A$ is the area of a transverse surface. The Hawking quasi-local energy
\cite{Haw} can be written as
\begin{equation}
E=\frac R{2G}\left(1+\frac1{16\pi}\oint{*}g(H,H)\right)
\end{equation}
where $G$ is Newton's constant. Since $g(H,H)=8\rho\rho'/\chi\bar\chi$, it can
be generalized to \cite{mon}
\begin{equation}
E=\frac R{4\pi G}\oint{*}\frac{K+\rho\rho'}{\chi\bar\chi}
\end{equation}
where
\begin{equation}\label{K}
K=\sigma\sigma'-\rho\rho'-\Psi_2+\Phi_{010'1'}+\Pi
\end{equation}
is Penrose's complex curvature (4.14.20), such that the Gaussian curvature is
$(K+\bar K)/\chi\bar\chi$, and $\Pi=\chi\bar\chi\Lambda$ (4.11.7). Penrose's
complex generalization (4.14.42--43) of the Gauss-Bonnet theorem is
\begin{equation}
\oint{*}\frac{K}{\chi\bar\chi}=\oint{*}\frac{\bar
K}{\chi\bar\chi}=2\pi(1-\gamma)
\end{equation}
where $\gamma$ is the genus of the transverse surface.

The propagation equations for the Hawking energy can be found from the
compacted spin-coefficient equations (4.12.32), using
$(\thorn,\thorn'){*}1=-2({*}\rho,{*}\rho')$, as \cite{mon}
\begin{eqnarray}
DE&=&\frac R{4\pi G}\left[\oint{*}\rho
\frac{K+\rho\rho'-\tau'\bar\tau'+\eth\tau'-\Phi_{101'0'}-3\Pi}{\chi\bar\chi}
\right.\nonumber\\
&&{}+\left.\oint{*}\rho'\frac{\sigma\bar\sigma+\Phi_{000'0'}}{\chi\bar\chi}
-\frac1A\oint{*}\rho\oint{*}\frac{K+\rho\rho'}{\chi\bar\chi}\right]\\
D'E&=&\frac R{4\pi G}\left[\oint{*}\rho'
\frac{K+\rho\rho'-\tau\bar\tau+\eth'\tau-\Phi_{010'1'}-3\Pi}{\chi\bar\chi}
\right.\nonumber\\
&&{}+\left.\oint{*}\rho\frac{\sigma'\bar\sigma'+\Phi_{111'1'}}{\chi\bar\chi}
-\frac1A\oint{*}\rho'\oint{*}\frac{K+\rho\rho'}{\chi\bar\chi}\right].\nonumber\\&&
\end{eqnarray}
For a uniformly expanding flow, $\rho$ and $\rho'$ may be taken outside the
surface integrals. One then sees that the terms in $K+\rho\rho'$ cancel, and
the terms in $\eth\tau'$ and $\eth'\tau$ integrate to zero by Penrose's version
of the Gauss divergence theorem (4.14.69), so that the energy propagation
equations simplify to
\begin{eqnarray}
DE&=&\frac R{4\pi G}\oint{*}
\left[\frac{\rho'(\sigma\bar\sigma+\Phi_{000'0'})}{\chi\bar\chi}\right.\nonumber\\
&&\qquad\qquad\quad
\left.{}-\frac{\rho(\tau'\bar\tau'+\Phi_{101'0'}+3\Pi)}{\chi\bar\chi}\right]
\label{prop1}\\
D'E&=&\frac R{4\pi G}\oint{*}
\left[\frac{\rho(\sigma'\bar\sigma'+\Phi_{111'1'})}{\chi\bar\chi}\right.\nonumber\\
&&\qquad\qquad\quad
\left.{}-\frac{\rho'(\tau\bar\tau+\Phi_{010'1'}+3\Pi)}{\chi\bar\chi}\right].
\label{prop2}
\end{eqnarray}

{\em Effective gravitational-radiation energy tensor.} Then one sees that the
non-matter terms can be encoded in a spinor $\Xi$ corresponding to $\Phi$ by
\begin{eqnarray}
\Xi_{000'0'}=|\sigma|^2\qquad&&\Xi_{010'1'}=|\tau|^2\nonumber\\
\Xi_{101'0'}=|\tau'|^2\qquad&&\Xi_{111'1'}=|\sigma'|^2
\end{eqnarray}
which can be written in a more manifestly invariant form as
\begin{eqnarray}
&&\Xi_{ABA'B'}=\gamma_{BE'DC}\bar\gamma_{B'ED'C'}\times\\
&&\quad(o_A\bar o_{A'}o^C\bar o^{C'}o^D\bar o^{D'}\iota^E\bar\iota^{E'}
+\iota_A\bar\iota_{A'}\iota^C\bar\iota^{C'}\iota^D\bar\iota^{D'}o^E\bar o^{E'})
\nonumber
\end{eqnarray}
where $\gamma_{AA'BC}$ are the spin-coefficients (4.5.2). Then $\Xi$ can be
seen to be a spinor in the normal space, since it involves $(o,\iota)$ only in
the combinations $l=o\bar o$, $l'=\iota\bar\iota$, not $m=o\bar\iota$, $\bar
m=\iota\bar o$. By comparing with the Einstein tensor
$G_{ab}=-2\Phi_{ABA'B'}-6\Lambda\varepsilon_{AB}\bar\varepsilon_{A'B'}$
(4.6.24) and the Einstein equation $G_{ab}+6\Lambda g_{ab}=-8\pi GT_{ab}$
(4.6.30), the effective gravitational-radiation energy tensor $\Theta$ is given
by
\begin{equation}
4\pi G\Theta_{ab}=\Xi_{ABA'B'}.
\end{equation}
The positivity of the components of $\Theta$ indicate that the gravitational
radiation carries positive energy; more exactly, $\Theta$ satisfies the
dominant energy condition.

Introducing a non-standard but logical index correspondence
$a=(0,3,2,1)\leftrightarrow AA'=(00',01',10',11')$, the components
$\Theta_{00}=|\sigma|^2/4\pi G$ and $\Theta_{11}=|\sigma'|^2/4\pi G$ have the
same form as the energy densities of transverse gravitational radiation at past
and future null infinity respectively \cite{mon,PR}, taking the usual
convention that $l$ is future outward and $l'$ is future inward. On physical
grounds, one might expect the other components to be longitudinal gravitational
terms. As a first step, consider the initial data for the dual-null
initial-value problem, based on a transverse surface $S$ and the null
hypersurfaces $(\Sigma,\Sigma')$ generated from it in the $(l,l')$ directions.
From the form of the spin-coefficient equations (4.12.32) and Bianchi
identities (4.12.36--41), the free vacuum initial data can be prescribed as
$(\rho,\rho',\Psi)$ on $S$, $(\sigma,\tau')$ on $\Sigma$ and $(\sigma',\tau)$
on $\Sigma'$. Then it is natural to interpret the last two pairs as ingoing and
outgoing radiation profiles respectively.

Szekeres \cite{GR} introduced a gravitational compass, a tetrahedral
arrangement of four test particles joined by springs, to characterize the
physical meanings of components of the Weyl tensor by their effect via the
geodesic deviation equation. In terms of the Weyl spinor $\Psi$, this
simplifies to: $\Psi_0$ for ingoing transverse radiation, $\Psi_1$ for ingoing
longitudinal radiation, $\Psi_2$ for the quasi-Newtonian (Coulomb-like)
gravitational field, $\Psi_3$ for outgoing longitudinal radiation and $\Psi_4$
for outgoing transverse radiation. A comparison can be made using the recent
unified framework for null and spatial infinity in terms of advanced and
retarded conformal factors $(\omega,\omega')$ \cite{inf}. For future null
infinity, one can use a radial null coordinate $r=\omega^{-1}$ such that
$\thorn r=1/\sqrt2$, $\thorn'r=0$. Quoting the asymptotic behaviour
$\rho\sim-1/\sqrt2r$, $\rho'\sim1/\sqrt2r$, $\sigma\sim\sigma_0r^{-2}$,
$\tau\sim\tau_0r^{-2}$, $\tau'\sim\tau'_0r^{-2}$, $\sigma'\sim\sigma'_0r^{-1}$
and assuming the standard ``peeling'' behaviour
$\Psi_\ell\sim\Psi_\ell^0r^{\ell-5}$ (9.7.38), (4.12.32c) yields
$\tau_0=-\bar\tau'_0$. Then the primed versions of (4.12.32b,c) yield
$\thorn'\sigma'_0=\Psi_4^0$ and $\thorn'\tau_0=-\bar\Psi_3^0$. Thus a wavelike
profile of $\Psi_3$ or $\Psi_4$ in retarded time $x'$ will drive a wavelike
profile of $\tau$ or $\sigma'$ respectively. With similar results at past null
infinity, this corroborates the proposal that the radiation may be encoded in
the relevant spin-coefficient, specifically $\sigma$ for ingoing transverse
radiation, $\tau$ for outgoing longitudinal radiation, $\tau'$ for ingoing
longitudinal radiation and $\sigma'$ for outgoing transverse radiation. From a
physical viewpoint, since $(|\sigma|^2,|\tau|^2,|\tau'|^2,|\sigma'|^2)/4\pi G$
have been identified as energy densities, it is perhaps more correct in general
to encode the radiation in the spin-coefficient than the component of $\Psi$,
since one can say that the radiation exists if and only if the relevant energy
density is non-zero.

It should be noted that longitudinal gravitational radiation does not exist in
linearized Einstein gravity or at infinity, where the Bondi flux is entirely
determined by the transverse gravitational radiation. Using the above
expansions at future null infinity, the energy density $\Theta_{11}$ of
outgoing transverse radiation falls off as $1/r^2$, while the energy density
$\Theta_{01}$ of outgoing longitudinal radiation falls off as $1/r^4$. Thus
longitudinal gravitational radiation from distant astrophysical sources will be
much too weak to measure. On the other hand, the energy density is generally
non-zero at finite distance and does contribute to the irreducible mass of a
black hole \cite{bhd,AK}.

{\em Energy conservation.} The spin-components of the generating vector $\xi$
and its dual
\begin{equation}
\tau=\hat{*}\xi
\end{equation}
may be written as $\tau=\tau^{00'}l+\tau^{11'}l'$ and
$\xi=\tau^{00'}l-\tau^{11'}l'$. A canonical flow of time is generated by the
recently introduced vector \cite{bhd}
\begin{equation}
\chi=\hat{*}\chi^*\qquad\chi^*=-g^{-1}(dR).
\end{equation}
Note potential confusion between these vectors $(\chi,\tau)$ and the
spin-coefficients denoted by the same symbols, but the distinction should be
clear in context. Since $DR=-R\rho$ and $D'R=-R\rho'$ for uniformly expanding
flows, one finds $dR=-R(\rho n+\rho'n')$, $\chi^*=R(\rho'l+\rho
l')/\chi\bar\chi$ and $\chi=R(\rho'l-\rho l')/\chi\bar\chi$. Note also that
$\chi^*=RH/2$ and $\chi=RH^*/2$.

Writing the covariant derivative of the energy as $\nabla_\xi
E=\tau^{00'}DE-\tau^{11'}D'E$, the four bracketed terms coming from the energy
propagation equations (\ref{prop1})--(\ref{prop2}) may all be written in a
similar form:
\begin{eqnarray}
\nabla_\xi E&=&\frac1{4\pi G}\oint{*}\left[\tau^{00'}\chi^{00'}(\Phi_{000'0'}+\Xi_{000'0'})\right.\nonumber\\
&&\qquad\qquad{}+\tau^{00'}\chi^{11'}(\Phi_{101'0'}+3\Pi+\Xi_{101'0'})\nonumber\\
&&\qquad\qquad{}+\tau^{11'}\chi^{00'}(\Phi_{010'1'}+3\Pi+\Xi_{010'1'})\nonumber\\
&&\qquad\qquad\left.{}+\tau^{11'}\chi^{11'}(\Phi_{111'1'}+\Xi_{111'1'})\right].
\end{eqnarray}
Collecting terms in an invariant form,
\begin{eqnarray}
\nabla_\xi E&=&\frac1{4\pi G}\oint{*}(\Phi_{ABA'B'}
+3\Lambda\varepsilon_{AB}\bar\varepsilon_{A'B'}+\Xi_{ABA'B'})\nonumber\\
&&\qquad\qquad{}\times\chi^{AA'}\tau^{BB'}.
\end{eqnarray}
In tensorial notation,
\begin{equation}
\xi^a\nabla_aE=\oint{*}(T_{ab}+\Theta_{ab})\chi^{a}\tau^{b}.\label{law}
\end{equation}
This has the form of an energy conservation law, expressing the increase in
mass-energy $E$ due to the energy densities of matter and gravitational
radiation, with respect to the canonical time vector $\chi$ and the normal
vector $\tau$, scaled correspondingly to the generating vector $\xi$. It has
the same form as the recently discovered energy conservation law for black
holes \cite{bhd}.

In an untrapped region where $\rho<0$ and $\rho'>0$, $\chi$ is future causal,
while $\tau$ is future causal if $\xi$ is outward achronal. Then the energy
conservation law (\ref{law}) for matter satisfying the dominant energy
condition implies the monotonicity property $\nabla_\xi E\ge0$ \cite{mon,Mar}.
In the special case of an inverse mean-curvature flow $\xi=\eta$, one has
$\tau\propto\chi$ and the energy conservation law shows monotonicity of $E$
under just the weak energy condition \cite{Fra}.

{\em Conclusion.} An energy conservation law has been derived for uniformly
expanding flows, generalizing inverse mean-curvature flows. It expresses the
change in Hawking energy as a surface integral of the energy densities of
matter and gravitational radiation, with respect to a vector $\chi$ giving a
preferred flow of time. The gravitational radiation is described by an
effective energy tensor $\Theta$ in the space normal to the surfaces. The Bondi
energy-loss equation can be recovered as the limit at future null infinity, but
the energy conservation law for black-hole horizons \cite{bhd} is independent,
the expansion of (the marginal surfaces in) the horizon being generally
non-constant on the marginal surfaces. Obtaining the same form of effective
energy tensor and energy conservation law is remarkable.

The results hold locally for a flow of any causal type, generated from any
initial surface on which $(\rho,\rho')$ do not change sign. However, the
physical interpretation in terms of energy cannot be so generally valid, since
the Hawking energy in flat space-time vanishes only for metric spheres
\cite{mon}. The main intended application is to spatial hypersurfaces
containing one or more marginal surfaces. The flow gives a locally unique way
to propagate a marginal surface outwards in the hypersurface. The level
surfaces can presumably be found numerically by modifying existing codes which
find level surfaces of a normalized null expansion \cite{Sch}. The inverse
mean-curvature flow has been instrumental in recent proofs of the Penrose
inequality in the case of a time-symmetric hypersurface \cite{PI}. More
exactly, the flow is generalized to a weak flow which can jump over obstacles
such as other black holes. If this can be implemented either analytically or
numerically for general uniformly expanding flows in, for instance, a binary
black-hole merger, the resulting weak flow could connect not only a common
black-hole horizon, but also individual black-hole horizons, to infinity.
Radiation profiles would be discontinuous at the jump, but such behaviour may
be expected due to gravitational lensing by the black holes themselves.

This energy conservation law therefore provides a measure of gravitational
radiation in the strong-field regime on and near a black hole, and perhaps all
the way out to infinity. In particular, the outgoing transverse radiation has
energy density $\Theta_{11}=|\sigma'|^2/4\pi G$. As a physical measure of the
waveform, one may take the component of energy density which contributes to
$D'E$ at infinity, $R\rho\Theta_{11}/|\chi|^2$, multiplied by $R^2$. The only
drawback of this proposal compared to the other made recently \cite{bhd} is
that one does not have exact advanced and retarded coordinates, since $(x,x')$
are defined separately at each moment in time. In practice, one could take
appropriate linear combinations of the area radius $R$ and the original time
coordinate. Then the waveform-prediction problem reduces to whether the
waveform, as a graph of $R^3\rho\Theta_{11}/|\chi|^2$ against retarded time at
given angle, converges in advanced time within the numerical domain.
Ironically, the structure required to define gravitational radiation is
provided by the black hole itself.

Thanks to Marc Mars, Walter Simon, Hugh Bray, Tom Ilmanen and Edward Malec for
discussions which clarified the nature of these flows and the generality of the
monotonicity of $E$, initially at the Erwin Schr\"odinger Institute, which
provided local support and hospitality. Research partly supported by NSF grants
PHY-0090091, PHY-0354932 and the Eberly research funds of Penn State.


\begin{references}
\bibitem{WT}J M Weisberg \& J H Taylor, {Phys. Rev. Lett.} {\bf 52}, 1348 (1984).
\bibitem{Is}R A Isaacson, {Phys. Rev.} {\bf 166}, 1263 \& 1272 (1968).
\bibitem{TGR}H Bondi, Nature {\bf 186}, 535 (1960);
H Bondi, M G J van der Burg \& A W K Metzner,
 {Proc. Roy. Soc. Lond.} {\bf A269}, 21 (1962);
R K Sachs, {Proc. Roy. Soc. Lond.} {\bf A270}, 103 (1962);
 {Phys. Rev.} {\bf 128}, 2851 (1962).
\bibitem{bhd}S A Hayward, {Phys. Rev.} {\bf 70}, 104027 (2004); {Phys. Rev. Lett.} {\bf 93}, 251101 (2004).
\bibitem{AK}A Ashtekar \& B Krishnan,
 {Phys. Rev. Lett.} {\bf 89}, 261101 (2002); {Phys. Rev.} {\bf D68}, 104030 (2003).
\bibitem{Haw}S W Hawking, {J. Math. Phys.} {\bf 9}, 598 (1968).
\bibitem{mon}S A Hayward, {Class. Quantum Grav.} {\bf 11}, 3037 (1994).
\bibitem{PI}H Bray, {J. Diff. Geom.} {\bf 59}, 177 (2001);
G Huisken \& T Ilmanen, {J. Diff. Geom.} {\bf 59}, 353 (2001).
\bibitem{Mar}M Mars, in {Proc. ``Mathematics of Gravitation II''} (2004),
www.impan.gov.pl/Gravitation/ConfProc.
\bibitem{HMS}S A Hayward, M Mars \& W Simon (in preparation).
\bibitem{PR}R Penrose \& W Rindler, Spinors and Space-Time Vols.\ 1 \& 2
 (Cambridge University Press 1984 \& 1988).
\bibitem{GR}P Szekeres, {J. Math. Phys.} {\bf 6}, 1387 (1965).
\bibitem{bhs}S A Hayward, {Class. Quantum Grav.} {\bf 11}, 3025 (1994).
\bibitem{Fra}J Frauendiener, {Phys. Rev. Lett.} {\bf 87}, 101101 (2001).
\bibitem{Tod}K P Tod, {Class. Quantum Grav.} {\bf 8}, L115 (1991).
\bibitem{inf}S A Hayward, {Phys. Rev.} {\bf D68}, 104015 (2003).
\bibitem{Sch}E Schnetter, {Class. Quantum Grav.} {\bf 20}, 4719 (2003).
\end{references}
\end{document}